\documentclass[12pt]{article}
\usepackage{graphicx}
\usepackage{amsmath}
\usepackage{amssymb}
\usepackage{latexsym}

\newcommand{\br}{\mathbb R}
\newcommand{\bz}{\mathbb Z}
\newcommand{\bc}{\mathbb C}

\newcommand{\brp}{{\mathbb R \mathbb P}^2}

\newcommand{\cD}{{\cal D}}

\newcommand{\ket}[1]{{|#1\rangle}{}}
\newcommand{\bra}[1]{{\langle#1|}}

\newcommand{\tq}{\tilde{q}}

\newcommand{\nn}{\nonumber\\}

\makeatletter
\@addtoreset{equation}{section}
\def\theequation{\thesection.\arabic{equation}}
\makeatother

\begin{document}

\vskip 7mm
\begin{titlepage}
 
 \renewcommand{\thefootnote}{\fnsymbol{footnote}}
 \font\csc=cmcsc10 scaled\magstep1
 {\baselineskip=14pt
 \rightline{
 \vbox{\hbox{hep-th/0203030}
       \hbox{UT-02-12}
       \hbox{March, 2002}
       }}}

 \vfill
 \baselineskip=20pt
 \begin{center}
 \centerline{\Large \bf 
   Crosscap States for Orientifolds of Euclidean $AdS_3$} 

 \vskip 2.0 truecm

\noindent{ \large Yasuaki Hikida}\footnote{
E-mail: hikida@hep-th.phys.s.u-tokyo.ac.jp}
\bigskip

 \vskip .6 truecm
 {\baselineskip=15pt
 {\it Department of Physics,  Faculty of Science, University of Tokyo \\
  Hongo 7-3-1, Bunkyo-ku, Tokyo 113-0033, Japan} 
 }
 \vskip .4 truecm

 \end{center}

 \vfill
 \vskip 0.5 truecm

\begin{abstract}
\baselineskip 18pt
Crosscap states for orientifolds of Euclidean $AdS_3$ are constructed.
We show that our crosscap states describe the same orientifolds which
 were obtained by the classical analysis. 
The spectral density of open strings in the system with
orientifold can be read from the M\"{o}bius strip amplitudes
and it is compared to that of the open strings stretched between branes
and their mirrors. We also compute the Klein bottle amplitudes.
\end{abstract}
 \vfill
 \vskip 0.5 truecm

\setcounter{footnote}{0}
\renewcommand{\thefootnote}{\arabic{footnote}}
\end{titlepage}

\newpage
 \baselineskip  18pt


\section{Introduction}
\indent

The string theories on $AdS_3$ have been much investigated in recent years
because they can be used to analyze the AdS/CFT
correspondence \cite{AdSCFT} beyond supergravity approximation.
These theories have non-trivial NSNS $H$-flux and can be described by the
$SL(2,\br)$ WZW models\footnote{The string theories on Euclidean $AdS_3$,
which will be studied in this paper, can be described by the $SL(2,\bc)/SU(2)$
WZW models.}.  
The closed string sector of these models has been studied for a decade.
The open string sector is now actively investigated  
by using the classical analysis \cite{Stan2,Stan3,BP,PR,LOPT} and
conformal field theory \cite{GKS,HS,RR,PS,Rajaraman,LOP,PST}.
The application to the AdS/CFT correspondence was also given in
\cite{BBDO}. 

The unoriented sector of the models can be obtained by including
orientifolds. 
In the terms of conformal field theory, D-branes are described by
boundary states \cite{ishibashi,cardy}.
Orientifolds can also be described by crosscap states
\cite{CLNY,Pol}, thus we can analyze the unoriented sector in the
similar way as the open string sector.  
Crosscap states in WZW models were investigated by algebraic
way \cite{sagnotti1,sagnotti2,sagnotti3} and recently geometric
aspects of crosscap states have been studied
\cite{oplane1,oplane2,oplane3,oplane4,pfoplane}. 

In this paper, we follow the analysis of boundary states in Euclidean 
$AdS_3$ \cite{GKS,PS,LOP,PST} which use the methods 
first developed in the context of Liouville field theory with
boundary \cite{FZZ,TLB,ZZ,Hos,PT,FH,ARS}. 
The information of boundary states can be given by one point
functions on the disk, however they are difficult to calculate directly. 
If we consider two point functions on the disk, we can obtain some
constraints on one point functions by comparing two different
expansions \cite{sewingbs} and one point functions are determined by
solving these constraints.
Now we are interested in the case of crosscap states and the same information
can be given by one point functions on worldsheet of
$\brp$. Just as the case of boundary states, we compare two
different expansions of two point functions and determine one
point functions by solving constraints we obtain \cite{sewingcs}.

The organization of this paper is as follows. 
In section 2, we review the closed string sector of string theories on
$AdS_3$ and discuss the geometry of D-branes and orientifolds.  
In section 3, boundary states are constructed by following the
analysis of \cite{LOP,PST}. 
In section 4, we obtain constraints of one point functions on
$\brp$ and find generic solutions. 
In section 5, we propose crosscap states for orientifolds with
the correct geometry.  Using the crosscap states, 
we compute Klein bottle and M\"{o}bius strip amplitudes.
From the information of the M\"{o}bius strip amplitudes, we
compare the spectral density of open strings in the system with
orientifolds to that of open strings
stretched between the mirror branes.
The conclusion and discussions are given in section 6.
In appendix A, we summarize the useful formulae.


\section{Review of String Theories on $AdS_3$}
\indent

The Lorentzian $AdS_3$ can be given by the hypersurface
as 
\begin{equation}
 (X_0)^2 - (X_1)^2 - (X_2)^2 + (X_3)^2 = L^2 ~,
\label{AdS(3)}
\end{equation} 
where $L$ is the radius of $AdS_3$ space and we will set $L=1$ for a while.
The Euclidean $AdS_3$ ($H_+^3$) can be obtained by the Wick rotation 
$X^3 = i X^3_E$. 
This space can be also realized by $SL(2,\bc)/SU(2)$ group manifolds
as 
\begin{equation}
 g = \left(
  \begin{array}{cc}
   \gamma \bar{\gamma} e^{\phi} + e^{-\phi} & - \gamma e^{\phi} \\
   - \bar{\gamma} e^{\phi} & e^{\phi}
  \end{array}
  \right)~,
\label{gamma}
\end{equation}
whose metric can be given by
\begin{equation}
 ds^2 = d \phi ^2 + e^{2 \phi} d \gamma d \bar{\gamma} ~. 
\end{equation} 
The coordinate $\phi$ describes the radial direction and $\phi \to
\infty$ corresponds to the boundary of Euclidean $AdS_3$, where
$\gamma$ and $\bar{\gamma}$ become the coordinates of the boundary. 

The string theory on this background is given by the
$SL(2,\bc)/SU(2)$ WZW models and they were well investigated\footnote{
See, for example, \cite{Gawedzki}.}.
The important class of functions on $H^3_+$ is given by
\begin{equation}
 \Phi_j (x,\bar{x},z,\bar{z}) = \frac{1 - 2j}{\pi} \left(
 \frac{1}{e^{-\phi} + | \gamma - x | ^2 e^{\phi} }
\right)^{2 j} ~,
\label{primary1}
\end{equation} 
which have the spin $j$ under the transformation of $SL(2,\bc)$ and 
$x ,\bar{x}$ labels some quantum numbers\footnote{
The labels $x$, $\bar{x}$ can be also identified as the boundary
coordinates in the sense of AdS/CFT correspondence \cite{AdSCFT}.
}.
The $SL(2,\bc)/SU(2)$ WZW models have conserved currents and primary
states.
The primary states correspond to the functions (\ref{primary1}) and transform
as
\begin{equation}
 J^a (z) \Phi_j (x,\bar{x},w,\bar{w}) \sim 
 - \frac{\cD^a}{z-w} \Phi_j (x,\bar{x},w,\bar{w}) ~.
\label{primary2}
\end{equation}  
Here $a  = \pm, 3$ and $\cD^a$ are given by
\begin{equation}
 \cD^+ = \frac{\partial}{\partial x} ~,~
 \cD^3 = x \frac{\partial}{\partial x} + j ~,~
 \cD^- = x^2 \frac{\partial}{\partial x} + 2jx ~.
\end{equation} 
The anti-holomorphic currents are defined in the same way.
The energy momentum tensor is given by Sugawara construction as
\begin{equation}
 T = \frac{1}{2(k-2)} ( J^+ J^- + J^- J^+ - 2 J^3 J^3 ) ~,
\end{equation}
where $k$ is the level of the models which is related to the radius $L$. 
The conformal weights of the primary fields
(\ref{primary1}) can be calculated by this energy momentum tensor as
\begin{equation}
 \Delta_j = - \frac{j(j-1)}{k-2}~. 
\end{equation}

The normalizable mode has the spin $j = 1/2 + is$, $s \in \br$ and
the Hilbert space of $SL(2,\bc)/SU(2)$ WZW models can be decomposed by
this label. Precisely speaking, this is the double counting because the
states with $j$ and $1-j$ are related as 
\begin{equation}
 \Phi_j (x, \bar{x}, z, \bar{z}) = 
  R(j) \frac{2 j - 1}{\pi} \int d^2 y | x - y |^{-4j} 
 \Phi_{1-j} (y, \bar{y}, z , \bar{z}) ~,
\label{RR}
\end{equation}
where
\begin{equation}
 R(j) = \nu^{1-2j} \frac{\Gamma (1 - u (2j -1))}{\Gamma (1 + u (2j -1))} 
  ~,~ \nu = \frac{\Gamma (1 - u)}{\Gamma (1+ u)} ~,
\end{equation}
and we use $u = 1 / (k-2)$.
Two point functions were calculated in \cite{Teschner} as
\begin{eqnarray}
 \lefteqn{\langle \Phi_{j} (x, \bar{x}, z , \bar{z})
          \Phi_{j'} (y, \bar{y}, w , \bar{w}) \rangle }\nn
  &=& \frac{1}{|z - w|^{4 \Delta_j} } \delta^{(2)} (x - y) 
    \delta (j + j' -1) 
  + \frac{B(j)}{|z - w|^{4 \Delta_{j}} |x - y |^{4j}}
  \delta (j - j') ~,
\end{eqnarray} 
where
\begin{equation}
 B(j) =\nu^{1-2j}  \frac{2j - 1}{\pi} 
 \frac{\Gamma (1 - u  (2j-1))}{\Gamma (1 + u (2j-1))} ~.
\end{equation}

In order to include branes and orientifolds, 
it is convenient to introduce other parametrizations of Lorentzian
$AdS_3$ space as 
\begin{equation}
 X_1 = \cosh \psi \sinh \omega ~, ~
 X_2 = \sinh \psi ~,~
 X_0 + i X_3 = \cosh \psi \cosh \omega e^{i t} ~,
\label{AdSc}
\end{equation}
where the metric is given by
\begin{equation}
 ds^2 = d \psi ^2 + \cosh ^2 \psi ( - \cosh^2 \omega dt^2 + d \omega ^2)~.
\end{equation}
The Euclidean version of $AdS_3$ space can be obtained by the Wick
rotation $t_E = i t$ just as before.

Maximally symmetric branes were
investigated in \cite{Stan2,Stan3,BP} classically and the geometry of
physical branes was identified as $AdS_2$ space, which corresponds to 
constant $\psi$ slice in the coordinates (\ref{AdSc}).
The open string stretched between two branes can be described by
worldsheet with boundary, and hence we have to assign boundary
conditions to currents. 
The boundary conditions for maximally symmetric branes are given in the
terms of boundary states
as\footnote{
The notation of currents is different from
that of \cite{HS}, so the same boundary conditions are given in the
different way.
}
\begin{equation}
 (J^a_n + \bar{J}^a_{-n}) \ket{B} = 0 ~.
\label{BS}
\end{equation}
In the next section, we will construct this type of boundary states by
following the analysis of \cite{LOP,PST}.

The geometry of the orientifolds was already discussed in \cite{oplane3}.
Orientifold operations are given by the combination of worldsheet
parity reversal ($\Omega : \sigma \to 2 \pi - \sigma$) and space
time $\bz_2$ isometries ($h =$diag$(\pm 1, \pm 1, \pm 1, \pm 1)$ in the
coordinates (\ref{AdS(3)})).
However, in order to preserve the non-trivial $H$-flux, we have to choose
$h$ which reverse the orientation of manifolds.
Moreover orientifolds must be time-like surfaces,
therefore we can only use $h = (+1, +1, -1, +1)$\footnote{
We can use the ones rotated by the symmetries. 
}.
This means $\psi = 0$ slice in the coordinates (\ref{AdSc}), 
thus the geometry of orientifolds is $AdS_2$ space. 
The corresponding crosscap states obey the conditions like boundary states as
\begin{equation}
 (J^a_n + (-1)^n \bar{J}^a_{-n}) \ket{C} = 0 ~.
\label{CS}
\end{equation}
In section 4 and 5, we will construct this type of crosscap states and
study their properties. 


\section{Boundary states for $AdS_2$ branes}
\indent

Boundary states can be constructed from the information of one point
functions on the disk with some boundary conditions.
The ansatz for one point function obeying the condition (\ref{BS}) was
proposed in \cite{LOP,PST} as
\begin{equation}
 \langle \Phi_j (x , \bar{x} , z, \bar{z}) \rangle _{\Theta}
 = \frac{U^{\pm}_{\Theta}(j) }{|x - \bar{x}|^{2j} |z-\bar{z}|^{2 \Delta_j}} ~,
\end{equation} 
where $+$ for $x_2 > 0$ and $-$ for $x_2 < 0$ ($x = x_1 + i x_2$). 
We have used $\Theta$ as the label of boundary conditions. 
The solution which obeys the boundary conditions (\ref{BS})
is locally given by $|x - \bar{x}|^{-2j}$.  
This solution has singular points along $Im x = 0$,
thus we can use the different ansatz across this line.
From the viewpoint of $AdS/CFT$ correspondence,
the $AdS_2$ branes can be domain walls to the boundary CFT at $Im x =0$,
therefore the discontinuity can be allowed.
However the coefficients $U^+_{\Theta}$ and $U^- _{\Theta}$ are not
independent but 
related by the reflection relations (\ref{RR}) as ($y = y_1 + i y_2$)
\begin{eqnarray}
 \lefteqn{\frac{U^{\pm}_{\Theta} (j)}{|x - \bar{x} | ^{2j}} = 
 R(j) \frac{2j -1}{\pi} \left(
 \int _{y_2 > 0} d^2 y \frac{ U^{+}_{\Theta} (1-j) |x - y |^{-4j}}{
     | y - \bar{y} |^{2(1-j)}} \right.} \nn && \hspace{5cm} +~ \left.
 \int _{y_2 < 0} d^2 y \frac{ U^{-}_{\Theta} (1-j) |x - y |^{-4j}}{
     | y - \bar{y} |^{2(1-j)} } 
\right) ~.
\label{UR} 
\end{eqnarray}
Integrating this equations, we obtain the following simple relations as
\begin{equation}
 U^{\pm} _{\Theta} (j) = R (j) U^{\mp} _{\Theta} (1-j) ~.
\end{equation}
Rewriting the coefficients as
\begin{equation}
 U^{\pm}_{\Theta} (j) = \Gamma (1 - u (2j - 1)) \nu^{1/2 - j} 
 f^{\pm}_{\Theta} (j) ~,
\end{equation}
we get the relations for $ f^{\pm}_{\Theta} (j)$ as 
\begin{equation}
 f^{\pm}_{\Theta} (j) = f^{\mp} _{\Theta} (1-j) ~.
\label{fpm}
\end{equation}

As we mentioned in the introduction, one point functions are difficult to
calculate and hence we utilize two point functions.
However general two point functions are also difficult to calculate,
therefore we make use of the state
$\Phi_{-\frac{1}{2}}$ belonging to the degenerate representation.
This state has the properties which make the analysis very simple as
\begin{equation}
 \partial^2 _x \Phi_{- \frac{1}{2}} (x,\bar{x},z,\bar{z}) = 0 ~,
\label{der}
\end{equation}
and hence the operator product expansions with $\Phi_j$ include only two
terms 
\begin{eqnarray}
 \lefteqn{
 \Phi_{-\frac{1}{2}} (x, \bar{x} ,z ,\bar{z}) 
 \Phi_j (y, \bar{y} ,w, \bar{w})
  \sim C_+ (j) |z-w|^{2u(1-j)} |x - y |^2 
  \Phi_{j+\frac{1}{2}}  (y, \bar{y} ,w, \bar{w})} \nn
  && \hspace{7cm} 
   + C_- (j) |z-w|^{2uj} \Phi_{j-\frac{1}{2}} (y, \bar{y} ,w, \bar{w}) ~.  
\label{OPE}
\end{eqnarray} 
The coefficients were obtained in \cite{Teschner} as
\begin{eqnarray}
 C_+ (j) &=&
  \nu \frac{\Gamma ( -u) \Gamma (1 +2u)}{\Gamma ( -2u) \Gamma ( 1 +u )} ~,\nn
 C_- (j) &=& 
   \frac{\Gamma ( -u) \Gamma ( 1 +2u ) \Gamma ( u(2j-2)) \Gamma (1-u(2j-1))}
   {\Gamma ( -2u) \Gamma ( 1 +u ) \Gamma ( u(2j-1)) \Gamma (1-u(2j-2))} ~.
\label{cpm} 
\end{eqnarray}
From these reasons we can calculate the following two point functions
\begin{equation}
 \langle \Phi_{-\frac{1}{2}} (x,\bar{x},z,\bar{z}) 
  \Phi_{j} (y,\bar{y},w,\bar{w}) 
 \rangle_{\Theta} ~,
\end{equation}
which will be used to obtain the constraints on one point functions.
If one state approaches to another state ($z \to w$, $x \to y$), 
it is natural to use the previous 
OPE (\ref{OPE}) and if the states become close to the boundary, it is
natural to expand by boundary operators.
Comparing the two expansions, we find the constrains as
\begin{equation}
 2 \sinh \Theta \cdot f^+_{\Theta} (j) =
  f^+ _{\Theta}(j + \textstyle \frac{1}{2}) 
 - f^+ _{\Theta}(j - \textstyle \frac{1}{2}) ~,
\end{equation}  
and general solutions can be given by the linear combinations of
\begin{equation}
 f^{\pm}_{\Theta} (j) = e^{\pm (2 \pi i n + \Theta) (2j -1)} ~,~
  e^{\pm (\pi i (2n+1) - \Theta) (2j -1)} ~,~ n \in \bz ~.
\end{equation} 

The authors \cite{LOP,PST} proposed the solutions which correspond to
the boundary states for $AdS_2$ branes as
\begin{equation}
 f^{\pm}_{\Theta} (j) = C e^{\pm \Theta (2j -1)} ~,
\label{BSS}
\end{equation}
where $C$ is some constant\footnote{
We will set $C=1$ because it does not affect the arguments below.
} independent of $j$.
The corresponding boundary states are given by
\begin{eqnarray}
 \lefteqn{\ket{\Theta}_C = 
  \int_{\frac{1}{2} + i \br_+} dj \left(
 \int _{x_2 > 0} d^2 x \frac{U^+_{\Theta} (1-j)}{|x - \bar{x}|^{2(1-j)}} 
  \ket{j,x,\bar{x}}_I \right.}\nn  && \hspace{4cm}+~\left.
  \int _{x_2 < 0} d^2 x \frac{U^-_{\Theta} (1-j)}{|x - \bar{x}|^{2(1-j)}} 
  \ket{j,x,\bar{x}}_I 
\right) ~,
\label{BS1}
\end{eqnarray}
where $\ket{j,x,\bar{x}}_I$ are ``Ishibashi'' boundary states based on the
primary states $\ket{j,x,\bar{x}}$. 

In fact, the terminology of ``Ishibashi'' boundary states is not accurate. 
The usual Ishibashi boundary states are defined by 
\begin{equation}
 \ket{j}_I = \sum_{I,J} M_{IJ}^{-1} J_{-I} \bar{J}_{-J} \ket{\Phi_j} ~,
\end{equation}
which obey the conditions (\ref{BS}) \cite{ishibashi,RR}. 
The labels $I$, $J$ are defined by the ordered 
set of $(a_i,n_i)$ $(n_i \geq 0)$ 
and $J_I = J^{a_1}_{n_1} \cdots J^{a^r}_{n_r}$. 
The coefficients $M_{IJ}$ are defined by 
$M_{IJ} = \bra{\Phi_j} J_I J_{-J} \ket{\Phi_j}$.
In our case, there is a discontinuity along $Im x = 0$, thus the
decomposition by the label $x, \bar{x}$ might be needed.
Then the  ``Ishibashi'' boundary states are defined by using the basis
$\ket{j,x,\bar{x}}$ and restricting the summation to non-zero modes as
\begin{equation}
 \ket{j,x,\bar{x}}_I = 
 \widetilde{\sum_{I,J}}  M_{IJ}^{-1} J_{-I} \bar{J}_{-J} \ket{j,x,\bar{x}}~.
\end{equation}

The geometry of the branes which the boundary states describe 
can be seen in the large $k$ limit by scattering with the
closed string states $\ket{g}$ which are localized at $g$ \cite{MMS} as
\begin{equation}
 \bra{g} j,x, \bar{x} \rangle = \Phi_j (x,\bar{x} | g) ~.
\end{equation}
The overlaps with the boundary states were calculated in \cite{LOP,PST}
and the results are given by
\begin{equation}
 \lim_{k \to \infty} \bra{g}\Theta \rangle _C = 
 \frac{1}{4\pi} \delta (\sinh \psi - \sinh \Theta ) ~,
\end{equation}
thus we can see that the boundary states (\ref{BS1}) describe $AdS_2$
branes at $\psi = \Theta$.

The annulus amplitudes can be given by the overlaps between two boundary
states as
\begin{eqnarray}
 \lefteqn{{}_C \bra{\Theta_1} 
 \tq^{\frac{1}{2}(L_0 + \bar{L}_0 - \frac{c}{12})} 
   \ket{\Theta_2}_C =
\int_{\frac{1}{2}+ i \br_+} dj \left(
 \int_{x_2 > 0} d^2 x 
 \frac{U^+_{\Theta_1}(j) U^+ _{\Theta_2} (1-j)}{|x-\bar{x}|^2} 
   \right.} \nn && \hspace{5.5cm} +~ \left.
 \int_{x_2 < 0} d^2 x 
 \frac{U^-_{\Theta_1}(j) U^-_{\Theta_2} (1-j)}{|x-\bar{x}|^2} 
\right) \frac{\tq^{u(j-\frac{1}{2})^2}}{\eta (\tq)^3} ~,
\label{AA}
\end{eqnarray} 
where $\tq = e^{2 \pi i ( -1/\tau )}$ 
is the closed string modulus and $c$ is the
central charge of the model. Using the modular transformation, the
amplitudes can be transformed into the open string channel. 
The modular transformation is given by
\begin{equation}
 \frac{s \tq^{us^2}}{\eta(\tq)^3 } = 2\sqrt{2u} \int ^{\infty} _0
 d s' \sin (4 \pi u s s')   \frac{s' q^{u{s'}^2}}{\eta(q)^3 } ~,~
 q = e^{2 \pi i \tau} ~,
\label{S-trans}
\end{equation}
where we use the modular transformation of $\eta$ function (\ref{etam}) 
and the annulus amplitudes can be rewritten as
\begin{eqnarray}
 \lefteqn{{}_C \bra{\Theta_1} 
 \tq^{\frac{1}{2}(L_0 + \bar{L}_0 - \frac{c}{12})} 
   \ket{\Theta_2}_C =}\nn
&& \int d^2 x \frac{ 1 }{| x - \bar{x} |^2} 
   \int ^{\infty} _0 d s' \frac{ \sqrt{2 u} \pi  \sinh (2 \pi s') s'}{
   \cosh (\frac{1}{u} (\Theta_1 - \Theta_2)) + \cosh(2 \pi s')}
   \frac{ q^{u{s'}^2}}{\eta (q)^3}~.
\end{eqnarray} 
The $x$ integral would be divergent, but this can be interpreted as the
divergence due to the integration of the infinite worldvolume of $AdS_2$
branes.  
The spectral density can be read from the annulus amplitude as
\begin{equation}
 \rho (s) \propto \frac{  \sinh (2 \pi s) s}{
   \cosh (\frac{1}{u} (\Theta_1 - \Theta_2)) + \cosh(2 \pi s)} ~,
\label{doso}
\end{equation}   
where the  spectrum belongs to the continuous series.
We should note that the coefficients are
not integers but non-negative real numbers, 
contrary to the rational conformal field theory case.


\section{Constraints for One Point Functions on $\brp$}
\indent

Orientifolds can be described by crosscap states, which can be
obtained by the information of one point functions on $\brp$. 
The ansatz for one point function can be given just like the
case of the boundary states as
\begin{equation}
 \langle \Phi_j (x,\bar{x},z,\bar{z}) \rangle _ {\brp} =
   \frac{U^{\pm}_{C} (j)}{|x-\bar{x}|^{2j} |1 + z \bar{z}|^{2 \Delta_j} }~. 
\end{equation} 
The $x$ dependence can be determined by the conditions (\ref{CS}) and
the ansatz of $+$ and $-$ are given for $x_2 >0$ and $x_2 < 0$,
respectively. 
The discontinuity across $Im x = 0$ exists for the same reasons as the
boundary states case.
The $z$ dependence can be obtained by the mirror technique
of unoriented worldsheet.
When we construct the boundary states for D-branes, we use the disk
amplitude, which is essentially identical to the amplitude on the upper half
plane. By using the mirror image technique, we can map the upper half
plane to the whole plane and vice versa. 
The reflection $I(z) = \bar{z}$ is used and it gives the fixed line at
$Im z = 0$.   
Now we are considering the unoriented worldsheet.
In this case, the reflection $I(z) = -1 / \bar{z}$ is used and the
worldsheet can be restricted to the upper half plane. 
This action gives no fixed line and the geometry becomes $\brp$.  
The one point function on the upper half plane can be given by the two
point function on the whole plane by making use of this reflection.

The analysis used in the previous section can be applied to the
case of orientifolds \cite{sewingcs}.
We again utilize the two point function on $\brp$ with
the primary $\Phi_{-\frac{1}{2}}$ as
\begin{equation}
 \langle \Phi_{-\frac{1}{2}} 
 (x,\bar{x},z,\bar{z}) \Phi_{j} (y,\bar{y},w,\bar{w}) 
 \rangle_{\brp} ~.
\label{tpf}
\end{equation}
Just as the case of the boundary states, we expand this quantity by two
different ways and we obtain the constraints by comparing the two
expansions. One way to express this quantity is to make use of the
operator product expansion (\ref{OPE}). 
This expansion is natural when the two primaries are close ($z
\to w$, $x \to y$) and it is given by\footnote{
We assume $Im y > 0$ and hence we use $U^+ (j)$. If we use $Im y < 0$,
we should replace $\pm$ by $\mp$ in the following discussions.
}
\begin{eqnarray}
 \lefteqn{ \langle \Phi_{-\frac{1}{2}} (x,\bar{x},z,\bar{z}) 
    \Phi_{j} (y,\bar{y},w,\bar{w}) \rangle_{\brp} = 
  \frac{|y - \bar{y} |^{-1-2j}}{|x - \bar{y}|^{-2}} 
  \frac{|1 + w \bar{w} |^{-\frac{3u}{2}-2\Delta_j}}{|1 + z \bar{w} |^{-3u}}}
  \nn && \hspace{3cm}
   \left(
    C_+ (j) U^+_C (j+ \textstyle \frac{1}{2}) F_+(\chi, \eta) +  
    C_- (j) U^+_C (j- \textstyle \frac{1}{2}) F_-(\chi, \eta)
   \right) ~,
\label{e1}
\end{eqnarray} 
where we have defined the cross ratios as
\begin{equation}
 \chi = \frac{|x - y |^2}{(x - \bar{x})(y - \bar{y})} ~,~
 \eta = \frac{|z-w|^2}{(1+z\bar{z})(1+w\bar{w})} ~.
\end{equation}
The functions
$F_{\pm} (\chi,\eta)$ 
are four point conformal blocks which were calculated in
\cite{Teschner} as
\begin{eqnarray}
F_+ (\chi,\eta) &=& \eta^{u(1-j)} (1-\eta)^{uj} \biggl(
 \chi F(-u,u+1,1-u(2j-1);\eta) \nn &&\hspace{3cm} +~ 
  \frac{uz}{1-u(2j-1)} F (1-u,u+1,2-u(2j-1);\eta)\biggr) ~,\nn
F_- (\chi,\eta) &=& \eta^{uj} (1-\eta)^{uj} \biggl(
 \frac{2j \chi}{1-2j} F(1+2uj,2u(j-1),1+u(2j-1);\eta) \nn &&\hspace{3cm} +~ 
  F (2uj,2u(j-1),u(2j-1);\eta) \biggr) ~.
\end{eqnarray} 
We use $F(a,b,c;\eta)$ as the hypergeometric functions whose
properties are summarized in appendix A.

The other way to express the two point function (\ref{tpf}) is to use
the operator product expansion between $\Phi_{-\frac{1}{2}}$ and 
the mirror image of $\Phi_j$. It is
natural to use this expansion when $z$ approaches to $-1/\bar{w}$ and $x$
approaches to $\bar{y}$, namely, when it can be expanded 
by $1-\eta$ and $1 - \chi$. 
Noticing $Im \bar{y} < 0$ from the assumption, we obtain
\begin{eqnarray}
 \lefteqn{ \langle \Phi_{-\frac{1}{2}} (x,\bar{x},z,\bar{z}) 
    \Phi_{j} (y,\bar{y},w,\bar{w}) \rangle_{\brp} = 
  \frac{|y - \bar{y} |^{-1-2j}}{|x - \bar{y}|^{-2}} 
  \frac{|1 + w \bar{w} |^{-\frac{3u}{2}-2\Delta_j}}{|1 + z \bar{w} |^{-3u}}}
  \nn && 
   \left(
    C_+ (j) U^-_C (j+ \textstyle \frac{1}{2}) F_+(1-\chi, 1-\eta) +  
    C_- (j) U^-_C (j- \textstyle \frac{1}{2}) F_-(1-\chi, 1-\eta)
   \right) ~.
\label{e2}
\end{eqnarray} 
Comparing two expansions (\ref{e1}) and (\ref{e2}), we get the
constraints of the coefficients $U^{\pm}_{\Theta} (j)$.
When comparing them, 
we use the following relations obtained by using the formula for
hypergeometric functions in appendix A as
\begin{eqnarray}
 F_+ (\chi,\eta) &=& \frac{\Gamma (1 - u (2j-1)) \Gamma (1 - u (2j -1))}{
           \Gamma (1 - u ( 2j -2)) \Gamma (1 - 2uj)} 
       F_- (1 - \chi, 1 - \eta) \nn
  && \hspace{1cm} -~ \frac{\Gamma (1 - u (2j-1)) \Gamma (u (2j -1))}{
           \Gamma (- u) \Gamma (1+u)} 
       F_+ (1 - \chi, 1 - \eta) ~,\nn
 F_- (\chi,\eta) &=&  \frac{\Gamma (u (2j-1)) \Gamma (u (2j -1))}{
           \Gamma (u ( 2j -2)) \Gamma (2uj)} 
       F_+ (1 - \chi, 1 - \eta) \nn
  && \hspace{1cm} -~ \frac{\Gamma (1 + u (2j-1)) \Gamma (- u (2j -1))}{
           \Gamma (- u) \Gamma (1+u)} 
       F_- (1 - \chi, 1 - \eta) ~.
\end{eqnarray}
Using the expressions of $C_{\pm} (j)$ (\ref{cpm}) and $f^{\pm}_C (j)$
defined by
\begin{equation}
 U^{\pm}_{C} (j) = \Gamma (1 - u (2j - 1)) \nu^{1/2 - j} f^{\pm}_{C} (j) ~,
\end{equation}
we obtain the following constraints as
\begin{eqnarray}
&&\hspace{0.7cm} f^+_C (j - \textstyle \frac{1}{2}) \sin (2 \pi u j) +
 f^+_C (j + \textstyle \frac{1}{2}) \sin (\pi u) =
 f^-_C (j + \textstyle \frac{1}{2}) \sin (\pi u (2j-1))~, \nn
&& f^+_C (j + \textstyle \frac{1}{2}) \sin (2 \pi u (j-1) ) -
 f^+_C (j - \textstyle \frac{1}{2}) \sin (\pi u) =
 f^-_C (j - \textstyle \frac{1}{2}) \sin (\pi u (2j -1))~.
\end{eqnarray}
General solutions of these equations are given by 
\begin{equation}
 f^{\pm}_C (j) = \pm C (j) 
  \cos (\pi u ( j - \textstyle \frac{1}{2})) ~,
\end{equation}
where $C (j)$ are sort of phase factors which satisfy
\begin{equation}
 C (j+1) = - C(j) ~,~ C(1-j) = - C(j) ~.
\label{PC}
\end{equation} 


\section{Crosscap States for Orientifolds}
\indent

The $AdS_2$ orientifolds are located on $\psi = 0$ in the coordinates 
(\ref{AdSc}) and hence we have to construct the crosscap states which
reproduce the geometry in the classical limit  ($k \to \infty$).
Therefore we propose the following solutions 
\begin{equation}
  U^+_C (j = \textstyle \frac{1}{2} + is ) = 
       \nu^{-is} \cosh ( \pi u s) \Gamma(1 -2ius)~,
\label{f+s}
\end{equation}
where we restrict the label $j$ to the normalizable mode. 
Because of this restriction we can use trivial phase factors which still
satisfy (\ref{PC}) along the shift of the real part of $j$.
At this stage, we can say at most that the solutions should be (\ref{f+s})
in the classical limit. However we will see below that the spectral
density of open strings in the system with orientifold reproduces that
of open strings stretched between the mirror branes. 
From these reasons we believe that our solutions are correct ones.
The crosscap states are constructed by these solutions as
\begin{eqnarray}
 \lefteqn{\ket{C}_C =  \int_{\frac{1}{2} + i \br_+} dj \left(
 \int _{x_2 > 0} d^2 x \frac{U^+_{C} (1-j)}{|x - \bar{x}|^{2(1-j)}} 
  \ket{C;j,x,\bar{x}}_I \right.}\nn  && \hspace{4cm}+~\left.
  \int _{x_2 < 0} d^2 x \frac{U^-_{C} (1-j)}{|x - \bar{x}|^{2(1-j)}} 
  \ket{C;j,x,\bar{x}}_I 
\right) ~,
\end{eqnarray}
where $\ket{C;j,x,\bar{x}}_I$ are ``Ishibashi'' crosscap states based on the
primary states $\ket{j,x,\bar{x}}$.  These states are defined just as
the ``Ishibashi'' boundary states.

The spectrum of closed strings in the system with orientifold can be
read from the Klein bottle amplitude. This amplitude can be obtained
from the overlap between two crosscap states and it is given by
\begin{eqnarray}
{}_C \bra{C} 
 \tq^{\frac{1}{2}(L_0 + \bar{L}_0 - \frac{c}{12})} 
   \ket{C}_C &=& \int_{\frac{1}{2}+ i \br_+} dj \left(
 \int_{x_2 > 0} d^2 x 
 \frac{U^+_{C}(j) U^+ _{C} (1-j)}{|x-\bar{x}|^2} 
   \right. \nn && \hspace{2cm} +~ \left.
 \int_{x_2 < 0} d^2 x 
 \frac{U^-_{C}(j) U^-_{C} (1-j)}{|x-\bar{x}|^2} 
\right) \frac{ \tq^{us^2}}{\eta (\tq)^3} \nn
&=& \int d^2 x \frac{1}{|x - \bar{x}|^2} \int^{\infty}_0 d s 
   \frac{\cosh ( \pi u s) 2 \pi u s}{\sinh (2 \pi u s)}
   \frac{ \tq^{us^2}}{\eta(\tq)^3 } ~.
\end{eqnarray} 
By using the modular transformation (\ref{S-trans}), we obtain
\begin{equation}
 {}_C \bra{C} 
 \tq^{\frac{1}{2}(L_0 + \bar{L}_0 - \frac{c}{12})} 
   \ket{C}_C =  \int d^2 x \frac{ 1 }{| x - \bar{x} |^2}
   \int ^{\infty} _0 d s' \frac{  \sqrt{2u} \pi s'}{
   \tanh (2 \pi s')}
   \frac{ q^{u{s'}^2}}{\eta (q)^3}~,
\end{equation} 
and the spectral density can be read as
\begin{equation}
 \rho (s) \propto  \frac{s}{
   \tanh (2 \pi s)} ~.
\end{equation}   
This quantity may be derived directly but it seems difficult since it
depends on the regularizations. 
Thus we will only compare below the spectral density of open strings
which can be easily compared with the spectral density previously
obtained (\ref{doso}). 

The spectrum of open strings in the presence of orientifold can be read
from the M\"{o}bius strip amplitudes. It is convenient to use the following
characters \cite{sagnotti1} in the calculation as
\begin{equation}
 \hat{\chi}_j ( q) = 
  e^{- \pi i (\Delta_j - \frac{c}{24})} \chi_j (- \sqrt{q})~,
\end{equation}
where we are using the characters
\begin{equation}
 \chi_s (q) = \frac{q^{us^2}}{\eta(q)^3 } ~.
\end{equation} 
The M\"{o}bius strip amplitudes are obtained as the overlaps between
boundary states and crosscap state as
\begin{eqnarray}
{}_C \bra{\Theta} 
 \tq^{\frac{1}{2}(L_0 + \bar{L}_0 - \frac{c}{12})} 
   \ket{C}_C &=& \int_{\frac{1}{2}+ i \br_+} dj \left(
 \int_{x_2 > 0} d^2 x 
 \frac{U^+_{\Theta} (j) U^+ _{C} (1-j)}{|x-\bar{x}|^2} 
   \right. \nn && \hspace{2cm} +~ \left.
 \int_{x_2 < 0} d^2 x 
 \frac{U^-_{\Theta}(j) U^-_{C} (1-j)}{|x-\bar{x}|^2} 
\right) \hat{\chi}_s (\tq)\nn
&=& \int d^2 x \frac{1}{|x - \bar{x}|^2} \int^{\infty}_0 d s
   \frac{\cosh ( \pi u s) \cos (2  \Theta s) 2 \pi u s}{\sinh (2 \pi u s)}
   \hat{\chi}_s (\tq) ~.
\end{eqnarray} 
In the case of M\"{o}bius strip amplitudes, the modular transformation
can be given by so-called  $P$ transformation ($P = T^{1/2} S T^2 S T^{1/2}$)
\cite{sagnotti1}. It transforms $\tau \to - 1 / (4
\tau)$ and in this case 
\begin{equation}
 \frac{s e^{2 \pi \tau u s^2}}{\eta(\tau)^3 } = 2\sqrt{u} \int ^{\infty} _0
 d s' \sin (2 \pi u s s')   
   \frac{s' e^{2 \pi (- 1 / 4 \tau) u{s'}^2}}{\eta(- \frac{1}{4\tau})^3 } ~.
\label{P-trans}
\end{equation}
Using this modular transformation, we obtain
\begin{equation}
 {}_C \bra{\Theta} 
 \tq^{\frac{1}{2}(L_0 + \bar{L}_0 - \frac{c}{12})} 
   \ket{C}_C =  \int d^2 x \frac{1 }{| x - \bar{x} |^2} 
   \int ^{\infty} _0 d s' \frac{(\sqrt{u} \pi /4) \sinh (2 \pi s')  s'}{
   \cosh (2 \pi s') + \cosh ( \frac{2}{u} \Theta)}
   \hat{\chi}_{s'} (q)~,
\label{MSA}
\end{equation} 
and the spectral density can be read as
\begin{equation}
 \rho (s) \propto  ~ \frac{\sinh (2 \pi s)  s}{
   \cosh (2 \pi s) + \cosh ( \frac{2}{u} \Theta)} ~.
\end{equation}   
This density is the same as that of
open strings stretched between the branes and their mirrors, namely, the
density (\ref{doso}) with $\Theta_1 = \Theta$ and $\Theta_2 = - \Theta$.

This correspondence is not accidental.
The M\"{o}bius strip amplitudes are reconstructed by the information of 
the annulus amplitudes and the behavior of the open string states under the
orientifold operation \cite{oplane3}.
As we said in section 2, the orientifold operation can be given by the
combination of worldsheet parity reversal $\Omega$ and space $\bz_2$
isometries $h$. 
In the coordinates (\ref{gamma}), $h$ acts as  
\begin{equation}
 h : ~ \phi \to \phi ~, ~ \gamma \to \bar{\gamma} ~,~ \bar{\gamma} \to \gamma
 ~,
\end{equation}
and for the boundary coordinates as $x \to \bar{x}$ and $\bar{x} \to x$.
Therefore the functions (\ref{primary1}) do not change under this
operation and hence the orientifold operation are expected to be
independent of $j$.  
Moreover the currents are transformed by the orientifold operation as
$J^a \to \bar{J}^a$. 
By following the analysis of \cite{oplane2} and using the above information, 
we can show that the M\"{o}bius strip amplitudes (\ref{MSA}) can be
correctly reconstructed by using the cylinder amplitudes between the
boundary states for the mirror branes (\ref{AA}). 
This is an attractive result and from this reason we can rely on our
choice of the solutions (\ref{f+s}).


\section{Conclusion}
\indent

We construct the crosscap states for the orientifolds of Euclidean
$AdS_3$ from the solutions of one point functions on $\brp$ (\ref{f+s}).
In the classical limit, we can show that these crosscap states describes 
the orientifolds with correct geometry.  
The Klein bottle and M\"{o}bius strip amplitudes are calculated and the open
string spectrum in the system with orientifold is compared to the
spectrum of open strings between D-branes and their mirrors.

We have to do more checks to obtain more evidences that our choice of
solutions is correct. One way is to make more constraints by using
other primaries. This seems to be very complicated but in principle we
can do. 
The other way is to compute the spectral density directly 
by other methods and to compare with ours. 
Since our orientifolds have infinite
volumes, we have to use some regularizations. In \cite{PST}, the open
string spectrum was derived directly and compared by using the cut-off
regularization. The leading terms are removed by using the
boundary states with reference boundary conditions $\Theta_{*}$.
Therefore in order to follow their methods in our case,
we might have to find the similar reference crosscap states.
In our paper, we closely follow the discussions in \cite{LOP} and
the spectral density is identified as the part which scales as the volume of
D-branes or orientifolds. However, the usual regularization might be
given by the cut-off regularization, thus it is important to study the
relation between these regularizations\footnote{
We are grateful to J.~Teschner for pointing out the regularization
dependence of the comparison of annuals and M\"{o}bius strip amplitudes.}.

Compared to the boundary states, many crosscap states are left to be  
constructed. For example, it would be interesting to construct the
crosscap states in Liouville theory or the orientifolds in $SU(N)$ WZW
models wrapping on the twisted conjugacy classes like ours.
It seems also important to apply to the AdS/CFT correspondence in
the system with orientifold. 


\section*{Acknowledgement}
\indent

We would like to thank Y. Sugawara and K. Hosomichi for useful
discussions.


\section*{Appendix A ~ Several Useful Formulae}
\setcounter{equation}{0}
\def\theequation{A.\arabic{equation}}
\indent

The hypergeometric functions have the following properties under the
reparametrizations 
\begin{eqnarray}
 F(a,b,c;z)&=& (1-z)^{c-a-b} F (c-a,c-b,c;z) ~,\\
 F(a,b,c;1-z)&=& \frac{\Gamma (c) \Gamma (c-a-b)}{\Gamma (c-a) \Gamma (c-b)}
  F (a,b,1-c+a+b;z) \nn
   &+& z^{c-a-b}   \frac{\Gamma (c) \Gamma (a+b-c)}{\Gamma (a) \Gamma (b)}
        F (c-a, c-b, 1+c-a-b;z) ~.
\end{eqnarray}
The Gauss recursion formulae for the parameters $(a,b,c)$ are given by
\begin{eqnarray}
&& c F (a,b,c;z) - (c-b) F (a,b,c+1;z) - b F (a,b+1,c+1;z) = 0 ~,\\
&&\hspace{-1cm} 
 c F (a,b,c;z) + (b-c) F (a+1,b,c+1;z) - b (1-z) F(a+1,b+1,c+1;z) = 0 ~,\\
&& c F (a,b,c;z) - c F (a+1,b,c;z) + b z F (a+1,b+1,c+1;z) = 0 ~. 
\end{eqnarray} 

We often use the following formulae for Gamma function as
\begin{eqnarray}
 \Gamma (1+z) &=& z \Gamma (z) ~,\\
 \Gamma (1-z) \Gamma (z) &=& \frac{\pi}{\sin(\pi z)}  ~,\\
 \Gamma (1 + ix) \Gamma (1 - ix)  &=& \frac{\pi x}{\sinh (\pi x)}~,
\end{eqnarray}
where $z$ is an arbitrary complex number and $x$ is a real number.

The Dedekind $\eta$ function is defined by
\begin{equation}
 \eta (\tau) = q^{\frac{1}{24}} \prod_{n=1}^{\infty} (1 - q^n) ~,
\end{equation}
where $q=\exp (2 \pi i \tau)$ and its modular transformation is given by
\begin{equation}
 \eta (\tau + 1) =
  e^{\pi i / 12} \eta (\tau) ~,~~
 \eta (- \frac{1}{\tau}) =
  \sqrt{-i \tau} \eta (\tau) ~.
\label{etam}
\end{equation} 



\end{document}